\documentclass[useAMS,amssymb,epsfig,usegraphicx,usenatbib,twocolumn]{emulateapj}

\def\araa{ARA\&A}
\def\apj{ApJ}
\def\apjl{ApJ}
\def\aap{A\&A}
\def\mnras{MNRAS}

\usepackage[usenames,dvips]{color}

\usepackage{epsfig}
\newif\ifAMStwofonts

\makeatother

\shorttitle{Survival of pure disk galaxies}
\shortauthors{Sonali Sachdeva and Kanak Saha}

\begin{document}
\title{Survival of pure disk galaxies over the last 8 billion years}

\author{Sonali Sachdeva and Kanak Saha}
\affil{Inter-University Centre for Astronomy and Astrophysics, Pune, India 411007}

\begin{abstract}

\noindent Pure disk galaxies without any bulge component, i.e., neither classical nor pseudo, seem to have escaped the affects of merger activity inherent to hierarchical galaxy formation models as well as strong internal secular evolution.

We discover that a significant fraction ($\sim 15 - 18\%$) of disk galaxies in the Hubble Deep Field ($0.4 < z < 1.0$) as well as in the local Universe ($0.02 < z < 0.05$) are such pure disk systems (hereafter, PDS). The spatial distribution of light in these PDS is well described by a single exponential function from the outskirts to the center and appears to have remained intact over the last 8 billion years keeping the mean central surface brightness and scale-length nearly constant. These two disk parameters of PDS are brighter and shorter, respectively, than of those disks which are part of disk galaxies with bulges.

Since the fraction of PDS as well as their profile defining parameters do not change, it indicates that these galaxies have not witnessed either major mergers or multiple minor mergers since $z\sim1$. However, there is substantial increase in their total stellar mass and total size over the same time range. This suggests that smooth accretion of cold gas via cosmic filaments is the most probable mode of their evolution. We speculate that PDS are dynamically hotter and cushioned in massive dark matter haloes which may prevent them from undergoing strong secular evolution.
\end{abstract}


\section{Introduction}
\label{sec:intro}
The $\Lambda$~Cold Dark Matter paradigm based on hierarchical structure formation \citep{WhiteRees1978} has shown remarkable success in explaining the large scale structure of our Universe. Since according to the models based on this paradigm, galaxies grow hierarchically, this in turn implies the inevitable role of major mergers and/or multiple minor mergers during their life-time \citep{Bournaudetal2007,Hopkinsetal2010,Naabetal2014}. Depending on the strength of these mergers, the final outcome could be scrambling of the preexisting disks, thickening of the disk structure \citep{TothOstriker1992} and formation of a classical bulge at the centre of the disks. These predictions appear to be in direct conflict to the findings from a number of surveys of the local Universe according to which there is an abundance of thin to super-thin disks, often without any merger-built classical bulges \citep{Kautsch2009,Weinzirletal2009,Kormendyetal2010,Butaetal2015}. These observations have posed one of the most formidable challenges to the current understanding of galaxy formation \citep{Baugh2006,Dutton2009,Scannapiecoetal2009,PeeblesNusser2010,Kormendy2015}.

Over the last couple of years, cosmological hydrodynamical simulations which use feedback due to supernovae or AGN and/or smooth, continuous accretion of cold gas via cosmic filaments have made tremendous progress towards making realistic disk galaxies \citep{Governatoetal2010,Agertzetal2011,Guedesetal2011,Stinsonetal2013,Marinaccietal2014}. However, the success is confined to reducing the bulge fraction to $\sim 30\%$ for moderate mass ($\sim 6\times10^{10}M_\odot$) galaxies. Disks more massive than this modest limit are constantly produced with significant bulge fraction.

Apart from the externally driven evolutionary causes, there is growing evidence that internally driven evolution of the disks also leads to the destruction of the pure exponential due to bulge formation \citep{Kormendy2015}. This formation is driven by non-axis-symmetric features (i.e., bars, spiral arms, etc.) which facilitate the transport of gas to the inner region \citep{KormendyKennicutt2004,Sahaetal2012,Athanassoula2012}. Whatever might have been the dominant evolutionary route (whether external or internal) - exponential stellar disks without any bulges (whether classical or pseudo) are not expected to be observed at the present epoch. 

Observationally, on the other hand, there is no dearth of such PDS galaxies in the local Universe \citep{Kormendyetal2010,Butaetal2015}. \citet{FisherDrory2011} finds that roughly $35\% \pm 12\%$ of the disk galaxies within the $11$~$Mpc$ volume are bulgeless. These observations suggest that PDS galaxies are not a rare phenomena rather might be the norm. In this paper, we search for the counterpart of this significant population in the high redshift Universe (till $z \sim 1$) using HDF observations. We trace the evolution of this population to find how they survived the merger violence and other disk instabilities to remain dynamically undisturbed. 

The paper is organized as follows: section~\ref{sec:data} describes briefly the data and analysis. Section~\ref{sec:modeling} describes identification and intrinsic properties of PDS and their observed structural parameters are detailed in section~\ref{sec:structure}. Discussion and primary conclusions are drawn in Section~\ref{sec:discussion}.

\begin{figure}
\rotatebox{0}{\includegraphics[height=6.2 cm]{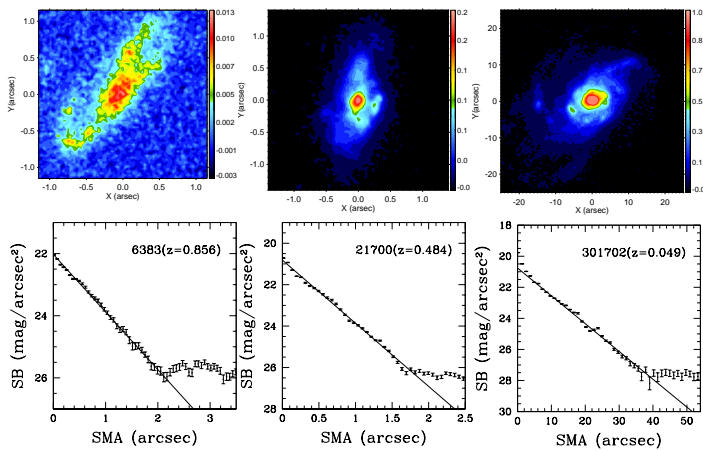}}
\caption{Top panel: Optical images of three PDS galaxies selected from three redshift ranges (mentioned above); the color bar represents the pixel intensity. Bottom Panel: The corresponding surface brightness profiles and fitted exponential function (solid lines). The magnitudes were calculated in the rest-frame {\it B} band.}
\end{figure}

\begin{figure*}[t]
\rotatebox{0}{\includegraphics[height=4.5 cm]{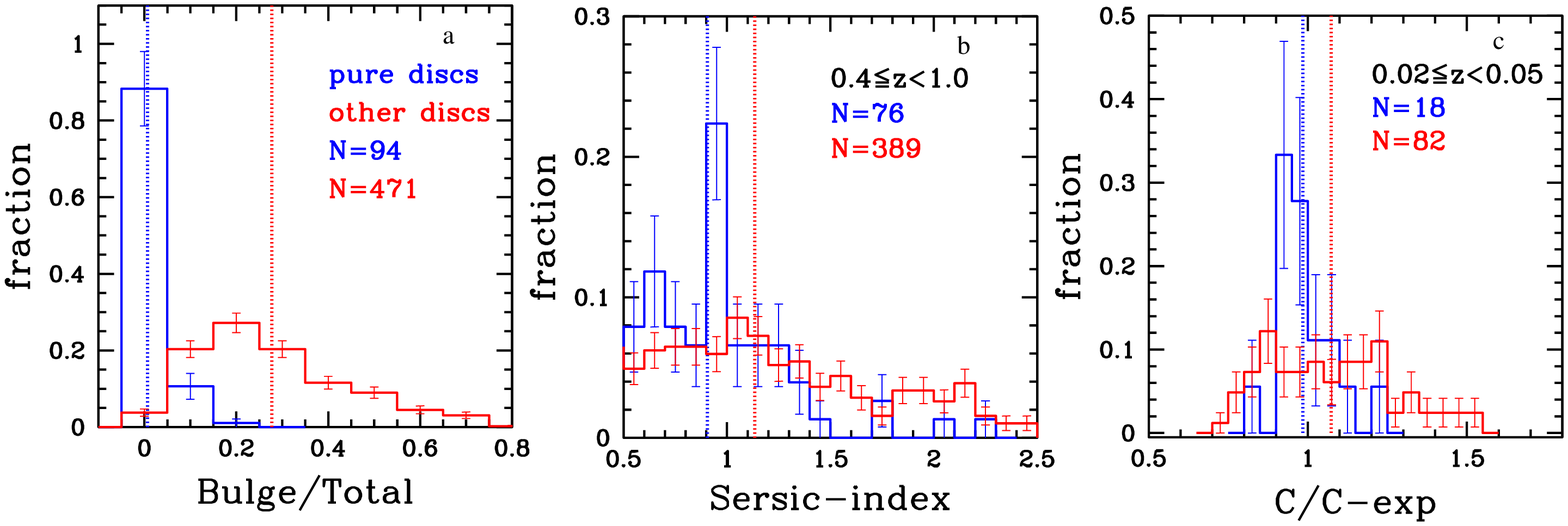}}
\caption{Panel a: showing the distribution of bulge-to-total (B/T) ratio for the selected 
PDS (in blue) and for the rest of the disk galaxies with bulges (red). Panels b, c: the distribution of global S\'ersic index (for $0.4<z<1.0$) and concentration index (for $0.02<z<0.05$) of the PDS (blue) in comparison to the rest of the disk galaxies (red) in the sample. Here, $C_{exp} = 2.87$ is the analytical concentration-index of a pure exponential disk.}
\end{figure*}

\section{Data}
\label{sec:data}
The full galaxy sample is same as that described in \citep{Sachdeva2013} (S13) and \citep{Sachdevaetal2015}. It overall comprises of $570$ disk-dominated (global S\'ersic-index less than $2.5$), bright (total {\it B}-band absolute-magnitude less than $-20$) galaxies in three redshift-ranges ($263$ in $0.77<z<1.0$, $203$ in $0.4<z<0.77$ and $101$ in $0.02<z<0.05$). The highest and middle redshift-range galaxy sample is from GOODS in CDF-S obtained using HST-ACS\footnote{Based on observations obtained with the NASA/ESA {\it Hubble Space Telescope}, which is operated by the Association of Universities for Research in Astronomy, Inc.(AURA) under NASA contract NAS 5-26555.} {\it V}, {\it i} and {\it z} filters \citep{Giavaliscoetal2004}. The redshifts for this sample are from the COMBO-17 survey \citep{Wolfetal2004}. The local galaxy sample is from NASA-Sloan Atlas\footnote{http://www.nsatlas.org. Funding for the NASA-Sloan Atlas has been provided by the NASA Astrophysics Data Analysis Program (08-ADP08-0072) and the NSF (AST-1211644).} based on SDSS \citep{Blantonetal2005}. The sample selection, completeness and image cleaning procedures are detailed in \cite{Sachdevaetal2015}.

Since the high redshift-range ($0.4-1.0$) images are from HST-ACS and local redshift-range ($0.02-0.05$) images are from SDSS, the image resolution in terms of physical galaxy size, is almost in the same range, i.e., $0.6-1.0$ kpc. Thus, the study has been carried out on resolution matched images.
   
Our selection of PDS galaxies is based on the isophotal analysis for which we have carried out
IRAF ellipse fitting task \citep{Jedrzejewski1987} on each galaxy image of the full disk-dominated
sample of $570$ galaxies. For each of these galaxies, we generated the intensity, ellipticity and position-angle profiles and visually inspected all of them to see if the profiles are well-behaved. We found 5 galaxies to have incognizable intensity profiles and, thus, were removed from the full sample. The remaining 565 galaxies were taken up for further analysis. 

To check for the robustness of radial intensity profiles in terms of the large redshift range of our study, we have used the FERENGI sample \citep{Bardenetal2008}. They provide local SDSS images of large morphological variety along with artifically redshifted images of the same set as they would appear from HST-ACS. We found the profiles to be as well defined both through visual inspection and quantitative profile measurement.

\section{Finding pure disk systems}
\label{sec:modeling}
A disk galaxy's total light is majorly contributed by two structural components - a stellar disk and a central bulge. The intensity profile of the disk part is described by an exponential function:

\begin{equation}
I_d(R) = I_o \exp(-R/R_d),
\end{equation}

\noindent where $I_o$ and $R_{d}$ are the central intensity and scale-length. Whereas, the intensity profile of the bulge part can best be described by a broad range S'ersic function:

\begin{equation}
I_{b}(R) = I_{e} \exp[-b_n((R/R_e)^{1/n}-1)],
\end{equation}

\noindent where $R_{e}$, $I_{e}$, and $n$ are the effective radius, intensity at the effective radius and the S\'ersic-index that determines the profile behavior. We fit the intensity profile of each galaxy individually with the combined disk + bulge intensity function i.e., $I_b(R) + I_d(R)$. The fitting has been accomplished using the non-linear-least-squares (NNLS) Levenberg-Marquardt algorithm. While fitting each profile individually, special care is given to visual inspection as well as the reduced chi-square value (note, no automated fitting is carried out on this sample). At the end of each fitting, we obtain five primary parameters that describe the disk and bulge contribution, and are taken for further calculation.

Based on the profile decomposition, we separate the sample of PDS galaxies. {\it As per definition, PDS are the ones whose intensity profile can be fitted well by a single exponential function ($I_d(R)$) all the way to the centre from the outskirts without requiring any bulge component}, see Fig.~1. These galaxies do not appear to harbour any bulge, neither classical or pseudo, as there is no 'extra-light' after accounting for the disk light. Of course, an $n=1$ pseudo-bulge could remain hidden in the galaxy, but it is unlikely that both pseudo-bulge component and the stellar disk would have the same central intensity and scale-length.

To ascertain our PDS selection, more quantitatively, we compute the values of disk-to-total ratio $D/T$ for all the galaxies. For a pure disk intensity profile, the luminosity within a radius R is defined as: 

\begin{equation}
\label{eq:diskL}
L_d(R) = 2 \pi I_{0} R_d^2 [1 - (1 + R/R_d)\exp(-R/R_d)].
\end{equation}

\noindent Then, analytically, for an ideal PDS (i.e., whose full light profile is a single smooth exponential), $D/T = 0.8$ (within $R = 3 R_d$). For our selected PDS sample, the average value comes out to be $\sim 0.73$ and for the rest of the disk galaxies in our sample, it is $\sim 0.44$. The results are very encouraging considering the fact that the small deviation from the expected theoretical value, is perhaps due to clumps that show up as bumps over the full profile. In addition to the $D/T$ ratio, we also computed the bulge-to-total ($B/T$) ratio for all the sample galaxies and its distribution is presented in Fig.~2(a). In accordance to expectations, for PDS the average $B/T$ is $0.014 \pm 0.030$, whereas for rest of the disk galaxies the average is $0.277 \pm 0.165$.

Both global S\'ersic index ($n$) and concentration index ($C$) are known to scale linearly with the $B/T$ ratio \citep{Grahametal2005,Gadotti2009}. Here, we check the distribution of these two parameters as well. The concentration index defined as $C = 5 \log[R_{80}/R_{20}]$ \citep{Blantonetal2003}, for an ideal smooth exponential profile, is $C_{exp} = 2.87$. We find that our PDS sample, in contrast to other disks, peaks at this value with a small scatter (Fig.~2(b)). Also, the global S\'ersic index value, obtained earlier in S13 by fitting the single S\'ersic component on all galaxy images, peaks at $n=1$ for the PDS sample with a small scatter (Fig.~2(c)). Thus, the analytical measurements are in excellent agreement with the profile fitting and visual selection.

\subsection{Population of pure disks}
\label{sec:fraction}
Out of $565$ disk galaxies, 94 i.e., $16.6\%$ are found to be PDS. The division is such that for the highest redshift range ($0.77 < z < 1.0$), it is 41 out of 263 i.e., $15.6\%$; for the middle redshift range ($0.4 < z < 0.77$), it is 35 out of 203 i.e., $17.2\%$; and for the local redshift range ($0.02 < z < 0.05$) it is 18 out of 101 i.e., $17.8\%$. It is interesting to note that the fraction of PDS galaxies over the last 8 billion years has not evolved much. And this brings back the puzzling question how they have survived so much merger violence as well as internal secular evolution. In the following section, we derive the structural parameters of these PDS which apparently survived the impact of the last 8 billion years of evolution.

\begin{figure*}
\label{fig:rprd}
\rotatebox{0}{\includegraphics[height=4.7 cm]{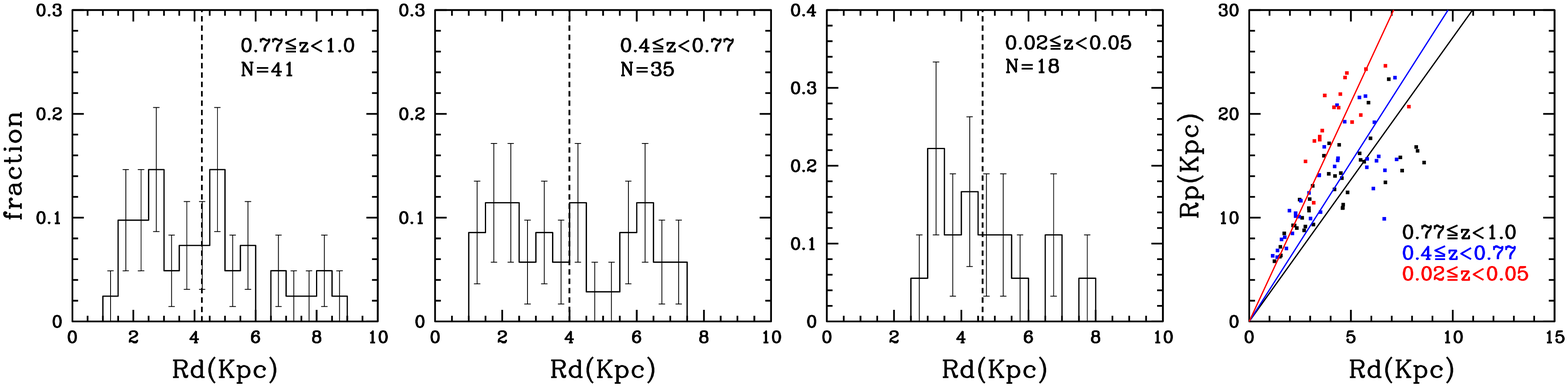}}
\caption{First three panels from left: The distribution of intrinsic scale length of PDS galaxies 
in the three redshift ranges. The solid line marks the mean at each redshift range. Extreme 
right panel: The Petrosian radius of PDS galaxies is plotted against their scale length. The solid lines mark the slope for a given redshift range.}
\end{figure*}

\section{Structural properties of pure disks}
\label{sec:structure}
For each galaxy in the PDS sample, we have the central intensity ($I_o$) and scale length ($R_d$). The total extent i.e., the Petrosian radius ($R_{p}$) was derived in \cite{Sachdevaetal2015} for the whole sample and we have used it here to compute the total luminosity using Eq.~\ref{eq:diskL}. Further, we use the redshift, current cosmological parameters and K-correction to obtain for each PDS galaxy: absolute magnitude, intrinsic central surface brightness in $mag/arcsec^2$ and scale-length in $kpc$ unit in rest frame {\it B}-band.

Fig.~3 shows the distribution of scale-length of the PDS in three redshift ranges. Clearly, these PDS show no preference to any particular value of $R_d$ rather support a wide range of scale-lengths irrespective of the redshift range. The average scale-length shows a statistically insignificant increase, as it goes from $4.2 \pm 0.31$ $kpc$ in the highest redshift range to $4.6 \pm 0.32$ $kpc$ for the local ones. However, the total size i.e., the Petrosian radius, in comparison, has undergone a significant increase of about $60\%$ since $z \sim 1$. To be precise, it increases from $12.7 \pm 0.31$ $kpc$ in the highest redshift range ($z \sim 0.77 - 1.$) to $20.5 \pm 1.11$ $kpc$ for the local ($z \sim 0.02 - 0.05$) disks. The extreme right panel of Fig.~3 shows the comparison between the scale-length and total size of the PDS galaxies. It is interesting to note that the total size of a PDS, on an average, is about $3 R_d$ in the highest redshift bin and has increased to about $4.5 - 5.0 R_d$ for the local ones, which is close to the Holmberg radius for many local disk galaxies. For disk galaxies with bulges, the increase in total size is even larger, i.e., $\sim87\%$ over the same time period, which is in agreement with the findings that late type galaxies undergo a highly substantial increase in size as they evolve with time \citep[e.g.,][]{Carrascoetal2010,vanDokkumetal2010,Cassataetal2013,McLureetal2013}. 

\begin{figure*}
\label{fig:SB}
\rotatebox{0}{\includegraphics[height=4.7 cm]{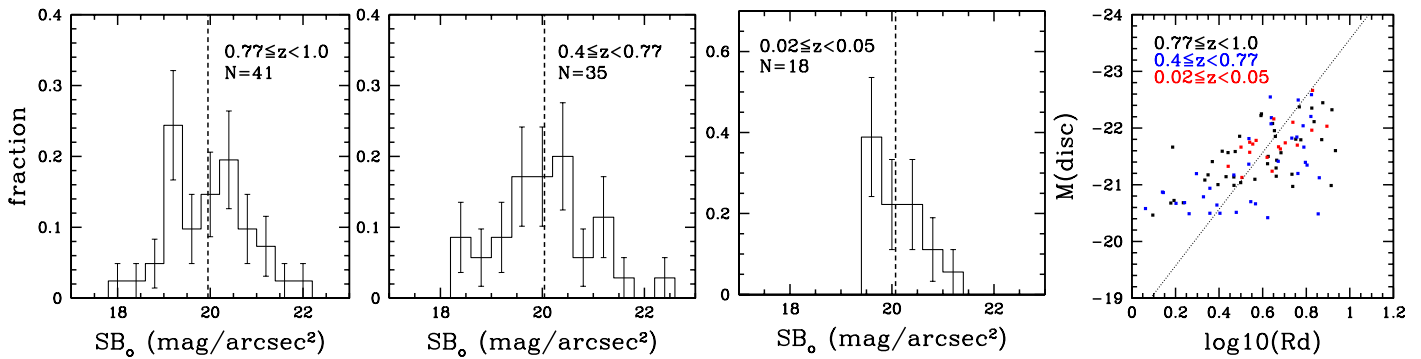}}
\caption{First three panels from left: same as in Fig.~3 but central surface 
brightness. Extreme right panel: The absolute magnitude of the PDS galaxies is plotted 
against their scale-length for the three redshift ranges. The solid line corresponds to 
$SB_{0} = 20 mag/arcsec^2$.}
\end{figure*}

The first three panels in Fig.~4 show the distribution of disk central surface brightness ($SB_{0}$) for the PDS in the three redshift ranges. The mean $SB_{0}$ remains nearly constant at $\sim 20 mag/arcsec^2$; with $19.95 \pm 0.14$ in ($0.77 \le z \le 1.$), $20.04 \pm 0.14$ in ($0.4 \le z \le 0.77$) and $20.07 \pm 0.11$ in ($0.02 \le z \le 0.05$). For comparison, we have also computed $SB_{0}$ of other disks, i.e., disks with pseudo and classical bulges. Note that the central surface brightness is of the disk component only and is obtained by the inward extrapolation of their disk intensity to the centre. The average values for the three redshift ranges are $21.30 \pm 0.07$, $21.47 \pm 0.08$ and $21.67 \pm 0.15 mag/arcsec^2$. These values are in remarkable agreement with the Freeman's law for the disk central surface brightness in bright local spiral galaxies \citep{Freeman1970}. It is intriguing that the observed $SB_{0}$ of disks with bulges are about 1.5 mag dimmer than those without bulges, i.e., the PDS. In the extreme right panel of Fig.~4, we show how absolute magnitude of the PDS varies with the disk scale length. For all redshifts, it is seen that more luminous disks have larger disk scale lengths, which is typically in agreement with the properties of other local exponential disks studied from SDSS \citep{Fathietal2010}. Overall, the PDS, in our current sample, seem to have undergone very little evolution both in terms of changes to the central surface brightness and the scale-length since redshift unity. 

Since the total size of the PDS galaxies has increased by about $60\%$, one would expect a corresponding growth in terms of stellar mass or luminosity of these galaxies. In the redshift range $z = 0.4 - 1.0$, we see that the stellar mass literally remains unchanged (with mean value nearly fixed at $\sim 6.57 \pm 1.19 \times 10^{10} M_{\odot}$) (computation of stellar mass is explained in \cite{Sachdevaetal2015}). However, a significant increase in stellar mass occurs from $z = 0.4$ to $0.02 - 0.05$, with mean mass going up to $\sim 9.29 \pm 1.49 \times 10^{10} M_{\odot}$ - amounting to about $41 \pm 2 \%$ change. This is in close agreement with what is known typically - that roughly half of the stellar mass of disk galaxies was assembled by $z=1$ \citep{Bundyetal2005,Mortlocketal2011,Marchesinietal2014,Ownsworthetal2014}. 

In addition, we computed the luminosity for these PDS and found that the average value goes from $3.18 \pm 0.31 \times 10^{10} L_{\odot}$ at the highest redshift range to $4.13 \pm 0.36 \times 10^{10} L_{\odot}$ at the local range, thus, increasing by $\sim 30 \pm 2 \%$. We note that the fractional increase in luminosity is lesser than that observed in it's stellar mass which is perhaps due to the fall in stellar-light/mass ratio with time \citep{Arnoutsetal2007,Droryetal2009,Pengetal2010}.

Simple calculation suggests that if a stellar disk (described well by an exponential light distribution) increases its total size from $3$ scale-length to $5$ scale-length, keeping its central surface brightness and scale-length intact, the luminosity increases by about $20\% $. Also, the extra light should augment in the outer parts of the disk without disturbing the slope of the surface brightness distribution. Going by this argument, we still have about $10 \%$ of luminosity increase and $20\%$ of stellar mass increase unaccounted for. We believe that this mismatch can largely be accounted for by relaxing the axis-symmetric assumption for the disks; the clumps in disks are asymmetric and show up as bumps in the intensity distribution over the exponential profile. 

\section{Discussion}
\label{sec:discussion}
The focus of this work has been to investigate the evolution of PDS i.e., disks without any bulge component, neither pseudo nor classical over the last $8$~billion years. We found that the fraction of these PDS has not altered much since $z \sim 1$; keeping it in the range of $15 - 18 \%$ in the three redshift bins studied here. In contrast, \cite{Sachdevaetal2015} found that amongst the rest of the galaxies in the sample, the fraction of disk galaxies with pseudo-bulges decreased with time, while that with classical bulges increased - suggesting some disk galaxies grew a classical bulge - which can be attributed to considerable minor mergers, gas clump migration to the centre, etc \citep{Elmegreenetal2008,Hopkinsetal2010, Weinzirletal2011,Buitragoetal2013, Naabetal2014}. It remains puzzling how these PDS held themselves away from such strong evolutionary forces. One of the possibilities is that the population of these PDS might be located in such secluded environment that interactions have not been substantial enough to have any affect on the morphology and thus, the fraction remains intact. But even without any environmental influence, internal secular evolution could bring substantial changes to the disk central surface brightness and scale-length and possibly lead to the formation of a pseudo-bulge at the centre \citep{KormendyKennicut2004}. What we found, on the contrary, is that both {\it disk scale-length and 
central surface brightness undergo statistically insignificant change from z$\sim$1 to the 
present epoch}. This reflects the remarkable robustness of the exponential profile once it is formed \citep{ElmegreenStruck2013}.

The PDS galaxies appear to be following a different evolutionary path compared to the disk galaxies with bulges (called the normal disk galaxies, which comprises the bulk of the disk galaxy population). We notice that both the disk central surface brightness and scale-length of normal disk galaxies are dimmer (by $\sim 1.5$ $mag$) and larger (by $\sim 1.5$ $kpc$), respectively, than those of the PDS at the present epoch. Although this suggests that PDS galaxies support distinct evolutionary histories, exact components which bring this distinction are not known, i.e., is it predominantly the environment or the feedback mechanism they sustain or the properties of the dark matter haloes they are embedded in. 

The intriguing part is that while the profile defining parameters remain intact, the total stellar mass and size of PDS increases substantially, $40\%$ and $60\%$ respectively, from $z \sim 1$ to $z\sim 0$. {\it Our analysis suggests that the size-mass evolution in the PDS galaxies occurred predominantly on the outskirts of the disk, keeping the slope of the intensity profile largely undisturbed}. In other words, the PDS galaxies evolved in such a way that nearly the "right" amount of mass got augmented in the periphery without altering the slope of the original exponential profile. Such an evolution could not have occurred if PDS galaxies were allowed to have undergone major mergers or multiple minor mergers. Perhaps, these PDS galaxies grew their mass and size through the smooth accretion of cold gas \citep{Lemoniasetal2011,Moffettetal2012,Moranetal2012} via cosmic filaments. Also, these galaxies might actually be located in the filaments connecting different clusters of galaxies.

We are thankful to the anonymous referee for careful reading and useful comments. 

\bibliographystyle{apj}

\end{document}